\documentclass[12pt]{article}
\usepackage{hhline}
\usepackage{latexsym,graphicx}
\usepackage{subfigure}
\usepackage{amssymb}
\usepackage{amsmath}
\usepackage{amscd}
\usepackage{amsthm}
\usepackage{float}
\usepackage{xcolor}
\usepackage[left=2cm,top=2.5cm,right=2.5cm,bottom=1.5cm]{geometry}   
 
\begin{document}
\begin{center}
\large{\bf{ The possibility of a stable flat dark energy-dominated Swiss-cheese Brane-world universe.}} \\
\vspace{10mm}
\normalsize{Nasr Ahmed$^1$,$^2$, Kazuharu Bamba$^3$ and F. Salama $^1$,$^4$}\\
\vspace{5mm}
\small{\footnotesize $^1$ Mathematics Department, Faculty of Science, Taibah University, Saudi Arabia.} \\
\small{\footnotesize $^2$ Astronomy Department, National Research Institute of Astronomy and Geophysics, Helwan, Cairo, Egypt\footnote{abualansar@gmail.com}} \\
\small{\footnotesize $^3$ Division of Human Support System, Faculty of Symbiotic Systems Science, Fukushima University, Fukushima 960-1296, Japan\footnote{bamba@sss.fukushima-u.ac.jp}} \\
\small{\footnotesize $^4$ Mathematics Department, Faculty of Science, Tanta University, Tanta, Egypt.\footnote{fatma.salama@science.tanta.edu.eg}}
\end{center}  
\date{}
\begin{abstract}
In this paper, we study the possibility of obtaining a stable flat dark energy-dominated universe in a good agreement with observations in the framework of Swiss-cheese Brane-world cosmology. Two different Brane-world cosmologies with black strings have been introduced for any cosmological constant $\Lambda$ using two empirical forms of the scale factor. In both models, we have performed a fine-tuning between the brane tension and the cosmological constant so that the EoS parameter $\omega(t)\rightarrow -1$ for the current epoch where the redshift $z\simeq 0$. We then used these fine-tuned values to calculate and plot all parameters and energy conditions. The deceleration-acceleration cosmic transition is allowed in both models, and the jerk parameter $j\rightarrow 1$ at late-times. Both solutions predict a future dark energy-dominated universe in which $\omega=-1$ with no crossing to the phantom divide line. While the pressure in the first solution is always negative, the second solution predicts a better behavior of cosmic pressure where the pressure is negative only in the late-time accelerating era but positive in the early-time decelerating era. Such a positive-to-negative transition in the evolution of pressure helps to explain the cosmic deceleration-acceleration transition. Since black strings have been proved to be unstable by some authors, this instability can actually reflect doubts on the stability of cosmological models with black strings (Swiss-cheese type brane-worlds cosmological models). For this reason, we have carefully investigated the stability through energy conditions and sound speed. Because of the presence of quadratic energy terms in Swiss-cheese type brane-world cosmology, we have tested the new nonlinear energy conditions in addition to the classical energy conditions. We have also found that a negative tension brane is not allowed in both models of the current work as the energy density will no longer be well defined.
\end{abstract}

\section{Introduction and motivation}

Numerous observations indicate that the expansion of the universe is accelerating \cite{11,13,14}. in order to explain such accelerated expansion, a mysterious form of energy with negative pressure has been proposed, dubbed as dark energy (DE), where the negative pressure acts as a repulsive gravity. Several dynamical scalar fields DE models have been suggested including quintessence \cite{quint}, Chaplygin
gas \cite{chap}, phantom energy \cite{phant}, k-essence \cite{ess}, tachyon \cite{tak}, holographic \cite{holog1, holog2} and ghost condensate \cite{ark,nass}. The accelerated expansion can also be investigated by modifying the geometrical part of the Einstein-Hilbert action \cite{moddd}, Examples include $f(R)$ gravity \cite{39} where $R$ is the Ricci scalar, Gauss-Bonnet gravity \cite{noj8}, $f(T)$ gravity \cite{torsion} where $T$ is the torsion scalar, and $f(R,T)$ gravity \cite{1}, where $T$ is the trace of the energy momentum tensor. For a general discussion of Dark Energy in the context of modified gravity see \cite{add1} where a detailed review has been given. A comprehensive report on $f(T)$ gravity has been given in \cite{add2} where various torsional constructions, from teleparallel, to Einstein-Cartan, and metric-affine gauge theories have been reviewed along with the corresponding cosmological applications. The relation with other modified gravity theories, such as $f(R)$ gravity, has also been considered. A very useful discussion for $f(R)$ gravity, in particular, and extended gravity, in general, has been presented in \cite{add3} where the future perspectives of extended gravity have also been considered. \par

Inspired by branes in string theory, brane-world models \cite{brane} have gained much attention as an interesting extra-dimensional alternative theory of gravity. In such models, our universe is a $3 + 1$ dimensional surface (brane) embedded in a $3 + 1 + d$ dimensional space-time called the bulk. While all particles and fields are trapped on the lower dimensional brane, only gravity leaks into the higher dimensional bulk. General Relativity (GR) is recovered at low energies where gravity is localized at the brane. Although the idea of the universe as a domain wall was first suggested in 1983 \cite{brane1}, a system of two branes of equal and opposite tension first received serious attention after the compactification of Horava-Witten theory to 5 dimensions \cite{brane2,nasrthesis}. in \cite{brane3}, the effective gravitational equations on the brane which can be reduced to Einstein equations in the low energy limit have been derived. It has been also shown that the brane cosmological constant depends on the brane tension and the bulk cosmological constant, this makes a fine-tuning necessary to get viable solutions. The modified Einstein equations on the brane can be written as:
\begin{equation}
G_{ab}=-\Lambda g_{ab}+\kappa^2T_{ab}+\tilde{\kappa}^4S_{ab}-\varepsilon_{ab},
\end{equation}
$G_{ab}$ is the Einstein tensor, $g_{ab}$ is the metric tensor and $\varepsilon_{ab}$ is the electric part of the Weyl curvature of the bulk. $S_{ab}$ is a quadratic expression in the energy momentum tensor $T_{ab}$:
\begin{equation}
S_{ab}=\frac{1}{12}TT_{ab}-\frac{1}{4}T_{ac}T^c_b+\frac{1}{24}g_{ab}(3T_{cd}T^{cd}-T^2),
\end{equation}
$\kappa^2$ and $\Lambda$ are the brane gravitational and cosmological constants respectively. They are related to the (positive) brane tension $\lambda$, the bulk cosmological constant $\tilde{\Lambda}$, and the bulk gravitational constant $\tilde{\kappa}^2$ by 
\begin{eqnarray}
6\kappa^2 &=&\tilde{\kappa}^2\lambda,\\
2\Lambda &=& \kappa^2\lambda+\tilde{\kappa}^2 \tilde{\Lambda}.
\end{eqnarray}
Fine-tuning is an unattractive feature in brane-world models where it is unlikely to have a relation between two independent quantities without a physical basis. As a solution to the long standing hierarchy problem, a specific brane-world set up has been proposed by Randall and Sundrum (RS model) in which the background metric is curved along the extra coordinate \cite{brane4, brane5}. This leads to an exponential suppression of energy scales between the two branes, and explains why the observed energy scales are so much smaller than the Planck scale. While the first RS model describes a system of two branes with positive and negative tensions, the second RS model describes only one brane with positive tension and infinite size extra dimension. There is also a fine-tuning between the brane tension and the bulk cosmological constant, the model becomes unstable under small deviations from this fine-tuning \cite{nasrthesis}. Negative tension branes are unstable objects, and it has been shown in \cite{brane6} that including a negative brane tension is unnecessary even in the case of 5D bulk. It has been also suggested that a variable brane tension can better describe our universe \cite{brane7}, in this case the sign of the brane tension may be changed according to the sign of the bulk cosmological constant \cite{brane6}.  \par
Several brane-world scienarios have been proposed such as the GRS model \cite{brane6i}, the DGP model \cite{brane6ii}, the thick brane model \cite{brane6iii}, and the universal extra dimension model \cite{brane6iiii}. Brane-world scienarios were also investigated in different modified
theories of gravity such as scalar-tensor gravity \cite{brane6iiii1}, metric $f(R)$ gravity \cite{brane6iiii2},
and $f(T)$ gravity \cite{brane6iiii3}. Brane-world cosmological models have several interesting features such as the self-acceleration of geometrical origin and the phantom behavior without a big rip singularity \cite{brane6iiii4}. In \cite{brane8}, a geometric alternative to the dark energy and cosmic acceleration has been proposed in the framework of brane-worlds where the extrinsic curvature of the brane is associated to the dark energy. The field equations of a brane-world model in modified $f(R, T)$ gravity have been derived in \cite{moreas}. The cosmological evolution with non-singular branes generated by a bulk scalar field coupled to gravity has been studied in \cite{brane9}. The effect of one extra dimension on the background cosmology and the dark energy properties has been investigated in \cite{brane10}. The idea of using a brane with variable tension to probe the late-time cosmic acceleration has been proposed in \cite{brane11}. The cosmic inflation driven by a dilaton-like gravitational field in the bulk has been studied in \cite{brane12}. A closed system of equations for scalar perturbations has been obtained in \cite{brane13} where the brane-world exhibits a phantom-like behavior at late times ($\omega<-1$) with no big-rip singularity (in contrast to conventional phantom models). A cosmological model which exhibits an early-time acceleration, middle-time deceleration and a late-time acceleration can provide a better description to cosmic evolution. For the sake of completeness, it is worthy to mention that late-time decelerating models can also be in a good agreement with observations and show a nice fit to some data \cite{sz1,sz2}. \par
In \cite{brane14}, an interesting brane-world cosmology with black strings (the simplest higher dimensional extensions of black holes) has been discussed, and the simplest scenario when black strings in the bulk penetrate the brane has been investigated. This penetration leads to the emergence of Schwarzschild black holes immersed in a Friedmann-Lemaitre-Robertson-Walker (FLRW) brane, and gives a Swiss-cheese structure to the brane. Such brane-world model may represent a better cosmological description where the Schwarzschild black holes have finite mass. Large scale structure for Swiss-Cheese model has been discussed in \cite{add4}. Some more advantages of the black strings have been mentioned in \cite{brane15}. In such braneworld scenario, the $4$D observable universe can be regarded as FLRW brane moving in a static $5$D Schwarzschild anti-de Sitter bulk. The FLRW metric given by
\begin{equation}
ds^{2}=-dt^{2}+a^{2}(t)\left[ \frac{dr^{2}}{1-Kr^2}+r^2d\theta^2+r^2\sin^2\theta d\phi^2 \right] \label{RW}
\end{equation} 
where $r$, $\theta$, $\phi$ are comoving spatial coordinates, $a(t)$ is the cosmic scale factor, $t$ is time, $K$ is either $0$, $-1$ or $+1$ for flat, open and closed universe respectively. We consider the flat case supported by observations \cite{flat1, flat2, flat3}, the spatially flat ($K=0$) cosmological evolution is governed by:
\begin{eqnarray}
\frac{\dot{a}^2}{a^2} &=& \frac{\Lambda}{3}+\frac{\kappa^2 \rho}{3}\left(1+\frac{\rho}{2\lambda}\right),  \label{cosm1}\\
\frac{\ddot{a}}{a} &=& \frac{\Lambda}{3}-\frac{\kappa^2 }{6}\left[\rho\left(1+\frac{2\rho}{\lambda}\right)+3p \left(1+\frac{\rho}{\lambda}\right) \right] .\label{cosm2}
\end{eqnarray}
Where $\Lambda$ is the cosmological constant, $\lambda$ is the brane tension, and GR can be recovered for $\rho/\lambda\rightarrow 0$. In the current work we are interested in a specific type of brane-worlds, namely the Swiss-cheese type brane-worlds. The absence of a stable dark energy dominated Swiss-cheese type braneworld universe in which a cosmic deceleration-acceleration transition can happen was the basic motivation behind the current work. In the original cosmological model for a brane-world with black strings \cite{brane14} there is no dark radiation, and the expansion of the brane-world universes forever decelerates regardless of the value of $\Lambda$. We have constructed two different models where, in contrast to the original model, we obtained dark energy-dominated universes with a deceleration-acceleration transition regardless of the value of $\Lambda$. While in the original Swiss-cheese brane-world cosmology \cite{brane14} the pressure is positive and the evolution can end in a 'pressure singularity', the behavior of cosmic pressure in the current work agrees with the dark energy assumption. The pressure in the first hyperbolic solution is always negative, but the second hybrid solution predicts a better behavior of cosmic pressure where it is negative only in the late-time accelerating epoch and positive in the early-time decelerating epoch.\par

Fine-tuning is one of the most controversial issues in modern physics and a major characteristic of brane-worlds. Another important feature of the current work is setting a dark energy observational restriction on performing the fine-tuning between the brane tension and the cosmological constant, we require that $\omega(t) \rightarrow -1$ for the current epoch where the redshift $z \rightarrow 0$. Using this observational restriction gives fine-tuned values to the brane tension, cosmological constant, and all free parameters in the model. This condition can always be used when performing fine-tuning in brane-world cosmology. Another remarkable feature of this work is that giving a negative value to the brane tension is not allowed because the energy density will no longer well defined. This result agrees with \cite{abed} where it has been shown that a negative tension brane is an unstable object. \par
Since black strings have been considered unstable and their instability has been discussed by many authors \cite{inst1,inst2}, this can reflect doubts on the stability of the associated cosmology. For this reason, we see that the stability of brane-worlds cosmological models with black strings (Swiss-cheese type brane-worlds) should be carefully investigated. Here, we have obtained a stable flat dark energy-dominated Swiss-cheese type brane-world cosmology in which the universe accelerates after an epoch of deceleration. Such a universe has not been obtained before in the literature for this specific type of brane-worlds.\par
The two functional forms of $a(t)$ used in the current work have been chosen for their consistency with observations according to the behavior of the deceleration and jerk parameters. In model $1$ we have used a hyperbolic form, and in model $2$ we have used a hybrid form. As we are goin to show in the next section, the deceleration parameters for both forms allow the cosmic deceleration-acceleration transition as they change sign from positive to negative. Also, the behavior of the jerk parameter for both forms shows that both models tend to flat $\Lambda$CDM for late-time where $j=1$. We have adopted an empirical approach in this work to solve the cosmological equations based on assuming an ansatz for $a(t)$ which allows the deceleration and jerk parameters to be consistent with observations. \par  

The rest of the paper is organized as follows: In section 2, we investigate the hyperbolic solution to the spatially flat cosmological equations and study the stability of the model. In section 3, we study the hybrid solution to the cosmological equations. The final conclusion is included in section 4.

\section{Model 1: Singular hyperbolic solution}
Since recent observations suggest a deceleration-acceleration transition \cite{11,rrr}, we can utilize an empirical form of the scale factor $a(t)$ which leads to a sign flipping of the deceleration parameter $q$ from positive (deceleration) to negative (acceleration). The following ansatz satisfies this condition for $0<n<1$ (Fig.1(a)):
\begin{equation} \label{ansatz}
a(t)= A \sinh^n(\xi t)
\end{equation} 
This hyperbolic empirical form has appeared in numerous settings of cosmology. It has been used in Bianchi cosmology and leaded to a good agreement with observations \cite{pr}. A Quintessence model has been introduced in \cite{sen} using $a(t)= \frac{a_o}{\alpha}[\sinh(t/t_o)]^{\beta}$ where $\beta$ is a constant, $\alpha = [\sinh(1)]^{\beta}$, $t_o$ and $R_o$ are the present time and the present day scale factor respectively. As has been noted in \cite{sen}, the main motivation behind using this hyperbolic form is its consistency with observations. In Chern-Simons modified gravity, the form $a(t)=\left(\frac{2\zeta}{3c_1}\right)^{\frac{1}{6}} \sinh^{\frac{1}{3}}(3\sqrt{c_1t})$ appears in the study of Ricci dark energy \cite{sent,sent11}. It has also been shown that the hyperbolic form $a(t)=\left(\frac{\Omega_m}{\Omega_{\Lambda}}\right)^{\frac{1}{3}} \sinh^{\frac{2}{3}}(\frac{3}{2}\sqrt{\Omega_{\Lambda}}H_ot)$ provides a unified analytic solution for cosmic evolution up to the late-time future \cite{sz}. The deceleration parameter is given as 
\begin{equation} \label{q1}
q(t)=-\frac{\ddot{a}a}{\dot{a}^2}=\frac{-\cosh^2(\xi t)+n}{\cosh^2(\xi t)},
\end{equation}
The current value of this parameter is expected to be in the neighborhood of $-0.55$ \cite{sz2}. The sign flipping is shown in Fig.1(a) for several values of $0<n<1$, in the current work we take $n=\frac{1}{2}$ where $q(t)$ varies in the range $-1 \leq q(t)\leq 1$. De-Sitter expansion occurs at $q=-1$, power-law expansion happens for $-1 < q < 0$, and a super-exponential expansion occurs for $q<-1$. Using (\ref{q1}), the cosmic deceleration-acceleration transition is expected to happen at some time where $q=0$ ( or $\ddot{a}=0$). For the current model we get
\begin{equation}
t_{q=0}=\frac{1}{2\xi}\ln(3+2\sqrt{2}),
\end{equation}
This gives $t\approx 0.88$ for $b=1$. The jerk parameter $j$ \cite{j1,j2} is given as 
\begin{equation}\label{jerk}
j=\frac{\dddot{a}}{aH^3}= 1+\frac{2n^2-3n}{\cosh^2(\xi t)},
\end{equation}
\begin{figure}[H]
  \centering    
	\subfigure[$q$]{\label{F636}\includegraphics[width=0.37\textwidth]{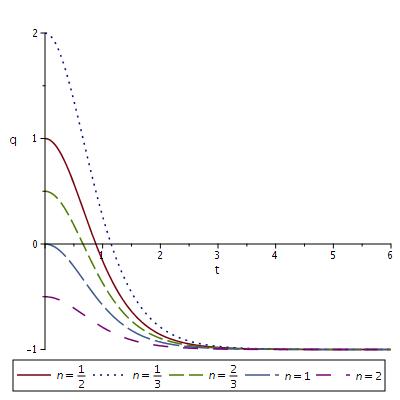}} 
	\subfigure[$j$]{\label{F63655}\includegraphics[width=0.37\textwidth]{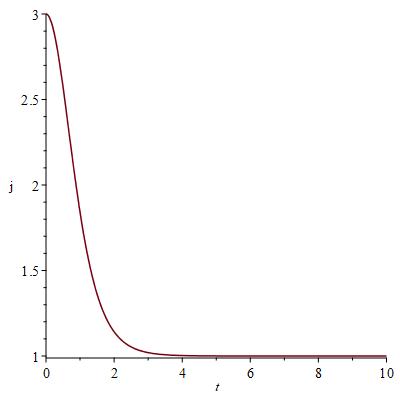}}
  \caption{  \ref{F636} The deceleration parameter changes sign from positive (decelerating phase) to negative (accelerating phase) according to observations for $0<n<1$. \ref{F63655} The jerk parameter $j=1$ at late-times where the current model tends to a flat $\Lambda CDM$ model. Any value in the range $0<n<1$ can be used, here we have taken $n=\frac{1}{2}$ and $\xi = 1$.}
  \label{fig:cassimir5cc5}
\end{figure}
where $\dddot{a}$ is the third derivative of $a(t)$. Since flat $\Lambda$CDM models have $j = 1$ \cite{j3}, we can describe models close to $\Lambda$CDM using the jerk parameter. Fig.1(b) shows that the current model tends to the flat $\Lambda$CDM model for late-time ($n=\frac{1}{2}$). Solving (\ref{cosm1}) and (\ref{cosm2}) with (\ref{ansatz}), the expressions for the pressure $p(t)$, energy density $\rho(t)$, and EoS parameter $\omega(t)$ can be written as 
\begin{equation}
\rho(t)= \frac{\sqrt{6\pi\lambda}\sqrt{(\xi^{2}+\frac{4\Lambda}{3}-\frac{16\pi\lambda}{3})e^{2\xi t}+\frac{1}{2}(\xi^{2}-\frac{4\Lambda}{3}+\frac{16\pi\lambda}{3})(e^{4\xi t}+1)}-4\pi \lambda(e^{2\xi t}-1)}{4\pi(e^{2\xi t}-1)}
\end{equation}
\begin{equation}
p(t)= \frac{\lambda}{16\pi}\frac{(16\pi\rho(t)-3\xi^2+4\Lambda)\cosh^2(\xi^2)-16\pi\rho(t)+2\xi^2-4\Lambda}{(\lambda+\rho(t))\sinh^2(\xi t)} ~~~~~~~~~~~~~~~~~~~~~~
\end{equation}
\begin{equation} \label{oo}
\omega(t)=  \frac{4p(t)\pi(e^{2\xi t}-1)}{\sqrt{6\pi\lambda}\sqrt{(\xi^{2}+\frac{4\Lambda}{3}-\frac{16\pi\lambda}{3})e^{2\xi t}+\frac{1}{2}(\xi^{2}-\frac{4\Lambda}{3}+\frac{16\pi\lambda}{3})(e^{4\xi t}+1)}-4\pi \lambda(e^{2\xi t}-1)}
\end{equation}
We need to fine-tune between the brane tension $\lambda$ and the cosmological constant $\Lambda$ to get the best agreement with observations. Since the equation of state parameter $\omega=-1$ for the current dark energy-dominated era with redshift $z\simeq 0$, the fine-tuning between $\lambda$ and $\Lambda$ should be performed such that $\omega=-1$ at $z\simeq0$. It can be easily seen that giving a negative value to the brane tension $\lambda$ is not allowed where the energy density $\rho(t)$ will never be well defined as there is a term $\sqrt{6\pi\lambda}$. This is a remarkable result where a negative tension brane is an unstable object \cite{abed}. It has been also shown in \cite{abed} that including a negative tension brane is unnecessary even in the 5D bulk case. Making use of the relation between the scale factor $a$ and the redshift $z$, $a=\frac{1}{1+z}$, we can express the cosmic time $t$ in terms of $z$ as $t=\frac{1}{\xi} \sinh^{-1}\frac{1}{A^2(1+z)^2}$. Substituting in (\ref{oo}) we directly get $\omega(z)$ as 
\begin{equation}\label{ww}
\omega(z)= \frac{1}{g(z)} \sqrt{\frac{8\pi}{3\lambda}}p(z)(e^{\frac{2}{\xi} \sinh^{-1}\frac{1}{A^2(1+z)^2}}-1)~~~~~~~~~~~~~~~~~~~~~~~~~~~~~~~~~~~
\end{equation}
where
\begin{eqnarray}
g(z)= \sqrt{(\xi^{2}+\frac{4\Lambda}{3}-\frac{16\pi\lambda}{3})e^{ \frac{2}{\xi} \sinh^{-1}\frac{1}{A^2(1+z)^2}}+\frac{1}{2}(\xi^{2}-\frac{4\Lambda}{3}+\frac{16\pi\lambda}{3})(e^{\frac{4}{\xi} \sinh^{-1}\frac{1}{A^2(1+z)^2}}+1)}\\   \nonumber
-4\pi \lambda(e^{\frac{2}{\xi} \sinh^{-1}\frac{1}{A^2(1+z)^2}}-1).~~~~~~~~~~~~~~~~~~~~~~~~~~~~~~~~~~~~~~~~~~~~~~~~~~~~~~~~~~~~~~~~~~~~~~~~
\end{eqnarray}
\begin{figure}[H]
  \centering            
  \subfigure[$\rho(t)$]{\label{F63}\includegraphics[width=0.35\textwidth]{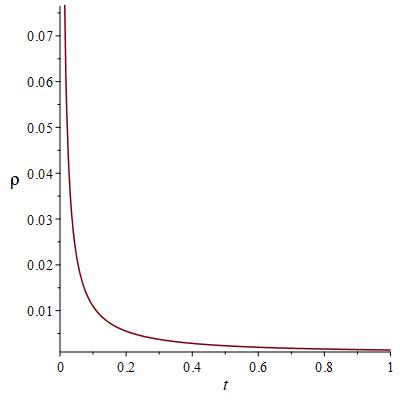}} 
		 \subfigure[$p(t)$]{\label{F6006}\includegraphics[width=0.35\textwidth]{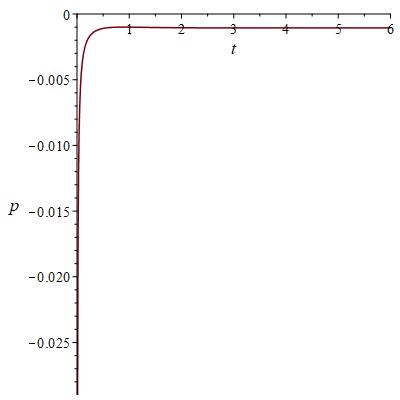}}\\
	 \subfigure[$\omega(t)$]{\label{F423}\includegraphics[width=0.35\textwidth]{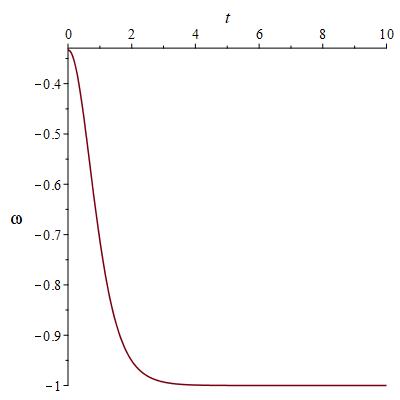}}  
	\subfigure[$\omega(z)$]{\label{F4273}\includegraphics[width=0.35\textwidth]{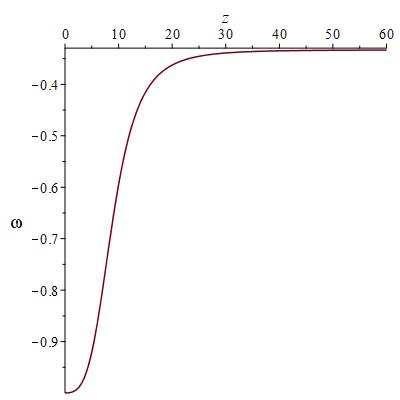}}
  \caption{The hyperbolic solution \ref{F63} Physically acceptable behavior of $\rho(t)$. \ref{F6006} $p(t)$ is always negative. \ref{F423} $\omega(t)$ lies in the quintessence region $ -1<\omega(t)< -\frac{1}{3}$. \ref{F4273} $\omega=-1$ at $z=0$. Here $\xi=1$, $\lambda=0.00002$, $\Lambda=0.02$ and $A=0.1$}
  \label{fig:cassimir55}
\end{figure}
Table $1$ shows that there is a wide range of the parameters satisfy $\omega(z=0)\simeq -1$. Setting $z=0$, $\lambda=0.00002$, $\Lambda=0.02$, $A=0.1$ and $\xi=1$ in (\ref{ww}) gives $\omega=-1$. Therefore, we are going to use these values to calculate and plot all parameters and energy conditions. The behavior of $p(t)$, $\rho(t)$ and $\omega(t)$ (Fig.2) has been found to be the same regardless the value of $\Lambda$, whether it is positive, negative, or zero. The physically acceptable behavior of the positive energy density is shown in Fig.2(a) where $\rho \rightarrow \infty$ as $t \rightarrow 0$. The negative pressure in Fig.2(b) supports the dark energy assumption where it represents a repulsive gravity effect which can accelerate the cosmic expansion. The EoS parameter lies in the quintessence region $ -1<\omega(t)< -\frac{1}{3}$ with no crossing to the cosmological constant boundary $\omega=-1$ i.e, no quintom behavior. A “no-go” theorem existing in quintom cosmology has been proved in \cite{nogo} in which the crossing of the EoS parameter of a single perfect fluid or a single scalar field to the phantom divide line at $\omega=-1$ is forbidden. Therefore, the no crossing of $\omega(t)$ to the phantom divide line in the current Swiss-cheese type brane-world model is expected.

\begin{table}[H]\label{tap}
\centering
\tiny
    \begin{tabular}{ | p{1cm} | p{0.8cm} | p{0.8cm} | p{0.8cm} | p{0.8cm} | p{0.8cm} | p{0.8cm} |p{0.8cm} | p{0.8cm} | p{0.8cm} | p{0.8cm} |}
    \hline
        A   & $0.1$ &  $2$  & 0.1 & $0.4$ & $0.6$ & 1& 3 & 0.7 &5&0.5\\ \hline
     $\xi$ & $1$ &  $2$ &  1 & $0.2$ & $3$ & 0.3 & 2 & 0.8 &5& 0.5\\ \hline
     $\lambda$ & $0.00002$ &  $0.00002$  &  0.02  & $0.3$ & $0.4$ & 1.4 & 5 & 1 &3& 0.1\\ \hline
   $\Lambda$  & $0.2$ & $0.2$ &   $0.1$ & $0.3$ & $0.6$ & 0.01 & 0.001& 1 &3 &0.1\\ \hline
	$\omega (z=0)$& $-0.9999$ & $-0.9465$ & $-0.9998$ & $-0.5087$ & $0.1391$ & $-0.9851$ & $-0.3486$ & $-1.6507$ & $-0.2954$ & $-0.9987$\\ \hline
    \end{tabular}
		\caption {Several values of $A$, $\xi$, $\lambda$ and $\Lambda$ can satisfy $\omega (z=0) \approx -1$.}
		\end{table}
		
\subsection{Stability of the model}
As we have mentioned in the introduction, the instability of black strings has been discussed in the literature by many authors \cite{inst1,inst2}. Since this instability can reflect doubts on the stability of brane-world cosmology with black strings, we now investigate the stability of the current model. The classical energy conditions (null, weak, strong and dominant) can be respectively written as \cite{hawk}: $\rho + p \geq 0$; $\rho \geq 0$, $\rho + p \geq 0$; $\rho + 3p \geq 0$ and $\rho \geq \left|p\right|$. Fig.3 (a) shows that only the dominant condition is not satisfied. In addition, because of the presence of quadratic energy terms in this model, we have also tested the new nonlinear energy conditions \cite{ec,FEC1, FEC2, detec, nasramri} which can be written as: (i) The flux energy condition (FEC): $\rho^2 \geq p_i^2$ \cite{FEC1, FEC2}. (ii) The determinant energy condition (DETEC): $ \rho\, .\, \Pi p_i \geq 0$ \cite{detec}. (ii) The trace-of-square energy condition (TOSEC): $\rho^2 + \sum p_i^2 \geq 0$ \cite{detec}. Fig.3 (b) shows that only the determinant energy condition is not satisfied.
\begin{figure}[H]
  \centering      
	\subfigure[Classical EC]{\label{F60016}\includegraphics[width=0.3\textwidth]{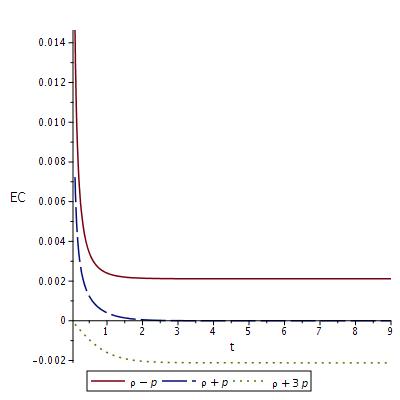}}
	 \subfigure[Nonlinear EC]{\label{F4213}\includegraphics[width=0.3\textwidth]{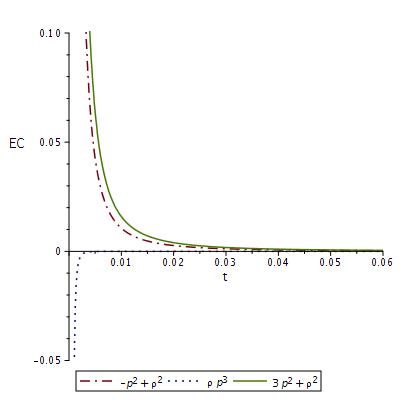}} 
		 \subfigure[$C_s^2$]{\label{F631}\includegraphics[width=0.3\textwidth]{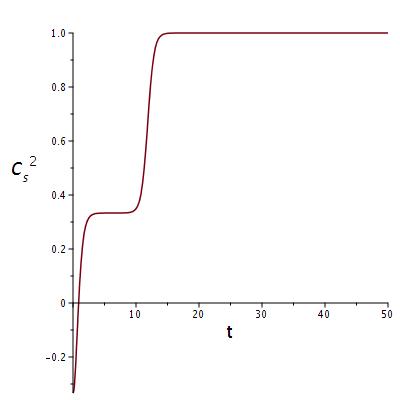}}
  \caption{ (a) and (b) The classical and nonlinear energy conditions plotted for: $\lambda=0.00002$, $\Lambda=0.02$, $A=0.1$ and $\xi=1$. (c) $0<C_s^2<1$ for the same values. }
  \label{fig:cassimir5jj5}
\end{figure}
The adiabatic square sound speed is given by: $c_s^2=\frac{dp}{d\rho}$. This quantity must be positive, and must satisfy the causality condition where the sound speed is less than the speed of light. In relativistic units where $c=G=1$, we can write $0 \leq \frac{dp}{d\rho} \leq 1$. For the current model we get, 
\begin{equation}
c_s^2=\frac{f(t)}{g(t)}~~~~~~~~~~~~~~~~~~~~~~~~~~~~~~~~~~~~~~~~~~~~~~~~~~~~~~~~~~~~~~~~~~~~~~~~~~~~~~~~~~~~~~~~~~~~~~~
\end{equation}
where $f(t)$ and $g(t)$ are given by
\begin{eqnarray}
f(t)= \left[\left((\xi^2-\frac{16}{3}\pi\lambda+\frac{4}{3}\Lambda)\cosh^4(\xi t) + (-\frac{5}{3}\xi^2-\frac{32}{3}\pi\lambda+\frac{8}{3}\Lambda)\cosh^2(\xi t)+\frac{1}{6}\sinh(\xi t) \right. \right.\\  \nonumber
\left.\left.  (\xi^2-16\pi\lambda+4\Lambda) \cosh(\xi t)+\frac{16}{3}\pi\lambda+\frac{2}{3} \xi^2-\frac{4}{3}\Lambda\right)(e^{2\xi t}+e^{4\xi t})\right.~~~~~~~~~~~~~~~~~~~~~~~~~~~~\\   \nonumber
\left. -\frac{1}{6}\sinh(\xi t)(e^{6\xi t}-1)\cosh(\xi t) \left(\xi^2+\frac{16\pi\lambda}{3}-\frac{4\Lambda}{3}\right)\right]e^{-2\xi t}(e^{2\xi t-1})^2.~~~~~~~~~~~~~~~~~~~~~~~
\end{eqnarray}
\begin{eqnarray}
g(t)= \sinh^4(\xi t)(e^{2\xi t+1})\left(\left(2\xi^2-\frac{32\pi\lambda}{3}+\frac{8\Lambda}{3}\right)e^{2\xi t}+(e^{4\xi t}+1)\left(\xi^2-\frac{16\pi\lambda}{3}-\frac{4\Lambda}{3}\right)\right).
\end{eqnarray}
Figure 3(c) shows that the condition $0 \leq \frac{dp}{d\rho} \leq 1$ is satisfied. 

\section{Model 2: Hybrid solution}
The hybrid scale factor is another empirical ansatz describes a deceleration-acceleration transition with a jerk parameter tends to the flat $\Lambda CDM$ model for late-time. It is given by \cite{hyb1, kumar}: 
\begin{equation} \label{hyb2}
a(t)=a_1 t^{\alpha_1} e^{\beta_1 t},
\end{equation}
where $a_1>0$, $\alpha_1 \geq 0$ and $\beta_1 \geq 0$ are constants. This ansatz is regarded as a generalization to power-law and exponential-law cosmologies where it is a mixture of both of them. For $\beta_1=0$ we obtain the power-law cosmology, and for $\beta_1=0$ we get the exponential-law cosmology. New cosmologies can be investigated for $\alpha_1>0$ and $\beta_1>0$. The deceleration parameter $q$ is given as 
\begin{equation}
q(t)=-\frac{\ddot{a}a}{\dot{a}^2}=\frac{\alpha_1}{(\beta_1t+\alpha_1)^2}-1
\end{equation}
The cosmic transition occurs at $t=\frac{\sqrt{\alpha_1}-\alpha_1}{\beta}$ which restricts $\alpha_1$ in the range $0<\alpha_1<1$ \cite{kumar}. The jerk parameter is given by
\begin{equation}
j={\frac {{\alpha_1}^{3}+ \left( 3\,\beta\,t-3 \right) {\alpha_1}^{2}+
 \left( 3\,{\beta}^{2}{t}^{2}-3\,\beta\,t+2 \right) \alpha_1+{\beta}^{3}
{t}^{3}}{ \left( \beta\,t+\alpha_1 \right) ^{3}}}
\end{equation}
Solving (\ref{cosm1}) and (\ref{cosm2}) with (\ref{hyb2}), we get
\begin{equation} \label{r}
\rho(t)= \frac{1}{2\pi\,t}\,\left(-2\,\pi\,\lambda\,t+\sqrt {-\pi\lambda\, \left(( -
4\,\pi\,\lambda\,-3\,{\beta}^{2}+\Lambda\,){t}^{2}-6\,
\alpha_1\,\beta\,t-3\,{\alpha_1}^{2} \right) }\right)
\end{equation}
\begin{equation} \label{p}
p(t)=\frac{1}{4}\,{\frac {\lambda\, \left( (4\,\pi\,\rho \left( t \right) -3
\,{\beta}^{2}+\Lambda\,)t^2-6\,\alpha_1\,\beta\,t-3\,{\alpha_1}^
{2}+\alpha_1 \right) }{\pi\,{t}^{2} \left( \lambda+\rho \left( t
 \right)  \right) }}
\end{equation}
The EoS parameter $\omega(t)=\frac{p(t)}{\rho(t)}$ is simply the division of (\ref{r}) and (\ref{p}). Following the same observational restriction we have set on the fine-tuning for the first hyperbolic solution ($\omega=-1$ at $z=0$), for this solution we take the following fine-tuned values: $\lambda=0.0009$, $\Lambda=0.037$, $\alpha_1=0.1$, $\beta=0.1$ and $a_0=1$. Therefore, we use these values in all calculations and plots. The behavior of $\rho(t)$, $p(t)$ and $\omega(t)$ (Fig.5) has been found to be the same regardless the value of $\Lambda$, i.e. whether $\Lambda$ is positive, negative or zero. Fig.5(a) shows the positivity of the energy density, it tends to $+\infty$ as $t \rightarrow 0$ as expected. Fig.5(b) shows that the pressure is positive during the early-time decelerating epoch and negative during the late-time accelerating epoch. This is a good agreement with the dark energy assumption where a component of negative pressure can push the universe to accelerate. This pressure behavior is better than the pressure behavior in the first hyperbolic solution where it is always negative through cosmic evolution. The EoS parameter lies in the region $ -1<\omega(t)< 2.3$ with no crossing to the cosmological constant boundary $\omega=-1$. 
\begin{figure}[H]
  \centering            
  \subfigure[$\rho(t)$]{\label{F639}\includegraphics[width=0.3\textwidth]{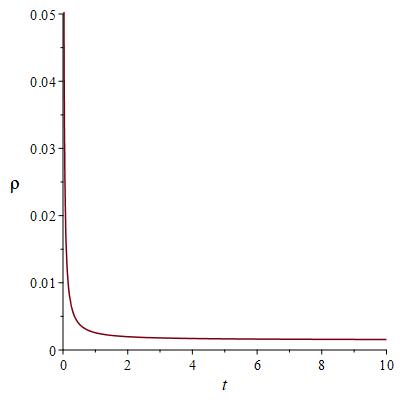}} 
		 \subfigure[$p(t)$]{\label{F60096}\includegraphics[width=0.3\textwidth]{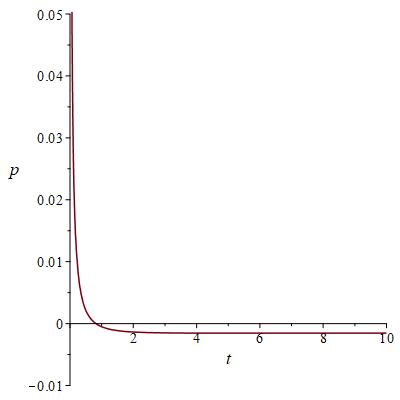}}
	 \subfigure[$\omega(t)$]{\label{F4293}\includegraphics[width=0.3\textwidth]{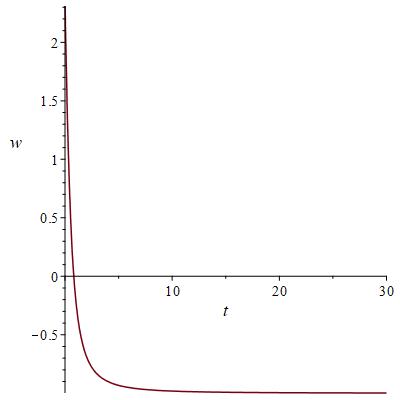}}  \\
	\subfigure[$\omega(z)$]{\label{F42793}\includegraphics[width=0.3\textwidth]{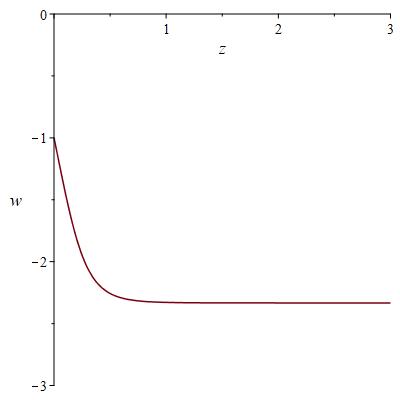}}
	\subfigure[$j$]{\label{F636595}\includegraphics[width=0.3\textwidth]{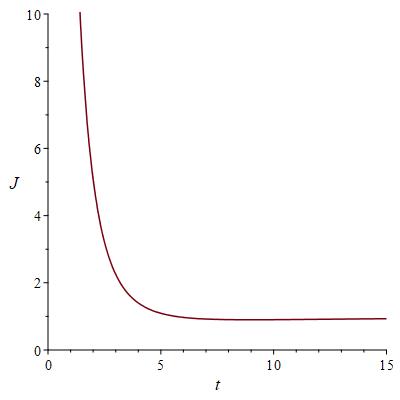}}
	  \subfigure[$q$]{\label{F6396}\includegraphics[width=0.3\textwidth]{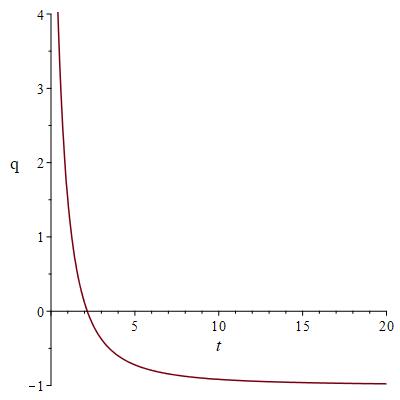}} 
  \caption{The hybrid solution \ref{F639} Positive energy density. \ref{F60096} The pressure is positive during the early-time decelerating epoch and negative during the late-time accelerating epoch. \ref{F4293} The EoS parameter lies in the region $-1<\omega(t)< 2.3$. \ref{F42793} $\omega=-1$ at $z=0$. \ref{F636595} The jerk parameter $j=1$ at late-times where the current model tends to a flat $\Lambda CDM$ model. \ref{F6396} The deceleration parameter flips sign from positive to negative according to observations. We have used the following fine-tuned values of this hybrid model: $\lambda=0.0009$, $\Lambda=0.037$, $\alpha_1=0.1$, $\beta=0.1$ and $a_0=1$.}
  \label{fig:cassi}
\end{figure}
\subsection{Non-Singular and cyclic solutions}
A non-singular solution of the form $a(t)= A\sqrt{\cosh(\xi t)}$ is found to be not allowed in the current model due to a nonphysical behavior of the positive energy density $\rho(t)$, we get $\rho \rightarrow 0$ as $t \rightarrow 0$ for any value of the cosmological constant $\Lambda$ and the brane tension $\lambda$. The same result has been obtained for the cyclic ansatz $
a(t)=A \exp \left[ \frac{2}{\sqrt{c^2-m^2}}\arctan \left(\frac{c\tan \left(\frac{kt}{2}\right)+m}{\sqrt{c^2-m^2}}\right)\right]$ where $A$ and $c$ are integration constants \cite{nasramri2,cuc}.
\section{Conclusion}

We have constructed two Swiss-cheese brane-world cosmological models with a deceleration-acceleration cosmic transition, the universe is flat and dark energy-dominated in both of them. This is in contrast to the original brane-world with black strings cosmological model \cite{brane14} where there is no dark radiation, and the universe forever expands and forever decelerates Regardless of the value of $\Lambda$. The major findings of the paper could be summarized in the following points:
\begin{itemize}
\item A fine-tuning between the brane tension and the cosmological constant has been performed based on observations where EoS parameter $\omega(t)\rightarrow -1$ when $z\simeq 0$ for the current epoch. The chosen fine-tuned values have been used in calculating and plotting the pressure, density, and stability conditions. 
\item The agreement of the model with observations is mainly represented in the deceleration-acceleration transition, the current dark energy-dominated epoch, and in the behavior of the jerk parameter for flat $\Lambda$CDM models where $j\rightarrow 1$. 
\item The no crossing to the cosmological constant boundary at $\omega=-1$ in the current work can be explained using a no-go theorem proved in \cite{nogo} where the crossing of the EoS parameter of a single perfect fluid or a single scalar field to the phantom divide line is forbidden. The modified cosmological equations $\ref{cosm1}$ and $\ref{cosm2}$ have been derived considering the perfect fluid matter as source in the universe.
\item While in the original Swiss-cheese Brane-world cosmology \cite{brane14} the pressure is positive and the evolution can end in a 'pressure singularity', the behavior of cosmic pressure in the current work agrees with the dark energy assumption. While the pressure in the first hyperbolic solution is always negative (repulsive gravity), the second hybrid solution predicts a better behavior of cosmic pressure where it is negative only in the late-time accelerating epoch but positive in the early-time decelerating epoch.
\item  Since the instability of black strings (which has been pointed out by some authors) might reflect doubts on the stability of cosmological models with black strings (Swiss-cheese type brane-worlds cosmological models), we have studied the stability through the sound speed, classical and non-linear energy conditions. A good level of stability has been obtained.
\item A negative tension brane is not allowed in the current work. This agrees with the result in \cite{abed} where it has been shown that a negative tension brane is an unstable object. 
\end{itemize} 
\section*{Acknowledgment}
We are so grateful to the reviewer for his many valuable suggestions and comments that significantly
improved the paper. The work of KB was partially supported by the JSPS KAKENHI Grant Number JP 25800136 and Competitive Research Funds for Fukushima University Faculty (19RI017).

\end{document}